\documentstyle[12pt]{article}

\def\be{\begin{equation}}
\def\ee{\end{equation}}
\def\bear{\begin{eqnarray}}
\def\eear{\end{eqnarray}}
\def\nn{\nonumber}

\def\hlf{{{1\over 2}}}

\newcommand\px[1]{{\partial_{#1}}}

\newcommand\inv[1]{{1\over{#1}}}

\def\eqdf{{\stackrel{def}{=}}}

\def\a{{\alpha}}
\def\b{{\beta}}
\def\g{{\gamma}}
\def\d{{\delta}}
\def\u{{\mu}}
\def\v{{\nu}}
\def\r{{\rho}}
\def\s{{\sigma}}
\def\t{{\tau}}
\def\h{{\eta}}

\def\const{{\mbox{const\ }}}

\def\cG{\widetilde{{\cal G}}}

\def\DD{{\Delta}}

\newcommand\WG[2]{{(\cG)^{#1}_{#2}}}            % Matrix of G
\newcommand\VG[2]{{(\cG^{-1})^{#1}_{#2}}}       % Inverse Matrix of G
\newcommand\GG[2]{{G^{#1}_{#2}}}                % Field strength  G
\newcommand\dG[2]{{\widetilde{G}^{#1}_{#2}}}    % Dual G
\newcommand\dGX[1]{{\widetilde{G}^{#1}}}        % Dual G

\newcommand\F[2]{{F^{#1}_{#2}}}                 % Field strength  F
\newcommand\dF[2]{{\widetilde{F}^{#1}_{#2}}}    % Dual F

\newcommand\bA[2]{{\overline{A}^{#1}_{#2}}}     % Barred gauge field A
\newcommand\bG[2]{{\overline{G}^{#1}_{#2}}}     % Barred Field strength  G

\newcommand\dl[1]{{\delta_{#1}}}                % Kronecker's
\newcommand\dlu[1]{{\delta^{#1}}}               % Kronecker's
\newcommand\ep[1]{{\epsilon_{#1}}}              % anti-symmetric tensor
\newcommand\epu[1]{{\epsilon^{#1}}}             % anti-symmetric tensor
\def\Lag{{\cal L}}                              % Lagrangian

\def\ra{{\bf \alpha}}                           % lattice root
\newcommand\tr[1]{{\mbox{tr}\{{#1}\}}}          % trace

\newcommand\rep[1]{{\underline{\bf {#1}}}}      % representation
\def\bM{{\bf M}}                                % matrix $M$
\def\bI{{\bf I}}                                % matrix $I$
\def\bfZ{{\bf Z}}                               % The integer

\def\hi{{\hat{i}}}                              % indices 1,2
\def\hj{{\hat{j}}}                              % indices 1,2
\def\hk{{\hat{k}}}                              % indices 1,2
\def\hl{{\hat{l}}}                              % indices 1,2
\def\Kv{{\nabla}}                  % Covariant (in curved space) derivative
\def\lam{{\lambda}}                % eigen-values of \bM

\newcommand\ZFUN[2]{{{\cal Z}^{({#1})}_{({#2})}}}    % partition function
\def\uv{{\mu\nu}}                                    % indices
\def\uva{{\mu\nu,a}}                                 % indices
\def\pu{{\partial_\mu}}                              % derivative
\def\pv{{\partial_\nu}}                              % derivative
\def\gh{{\hat{g}}}                                   % unit iso-vector

\def\DAu{{[{\cal D}A_\mu]}}                          % measure
\def\DFuv{{[{\cal D}F_{\uv}]}}                       % measure
\def\DGuv{{[{\cal D}G_{\uv}]}}                       % measure
\def\DGtuv{{[{\cal D}\widetilde G_{\uv}]}}
\def\Dgh{{[{\cal D}\gh]}}                            % measure
\newcommand\cD[1]{{[{\cal D}{#1}]}}                  % general measure

\def\as{{\alpha_s}}                                  % coupling
\def\fvas{{1\over{4\alpha_s}}}                       % 1/4 inverse coupling
\def\epabc{{\epsilon_{abc}}}                         % anti-symmetric
\def\epuv{{\epsilon^\uv}}                            % anti-symmetric

\newcommand\Fu[2]{{F^{{#2},{#1}}}}                   % Field strength  F
\newcommand\Fd[2]{{F^{#1}_{#2}}}                     % Field strength  F
\newcommand\Fud[3]{{{{F^{#2}}_{#3}}^{#1}}}           % Field strength  F
\newcommand\Fdu[3]{{{F_{#2}}^{{#3},{#1}}}}           % Field strength  F

\newcommand\Wd[1]{{W_{#1}}}                     % covariant W tensor
\newcommand\Wu[1]{{W^{#1}}}                     % contra-variant W tensor
\newcommand\Wud[2]{{{W^{#1}}_{#2}}}             % W tensor
\newcommand\Wdu[2]{{{W_{#1}}^{#2}}}             % W tensor

\def\SW{{V}}
\newcommand\SWd[1]{{\SW_{#1}}}                  % covariant hat W tensor
\newcommand\SWu[1]{{\SW^{#1}}}                  % contra-variant hat W tensor
\newcommand\SWud[2]{{{\SW^{#1}}_{#2}}}          % hat W tensor
\newcommand\Hd[1]{{h_{#1}}}                     % covariant h tensor
\def\WW{{(W^2)}}                                % trace of W^2
\newcommand\WWd[1]{{{(W^2)}_{#1}}}              % covariant W^2 tensor
\newcommand\Rd[1]{{R_{#1}}}                     % curvature tensor
\newcommand\pps[3]{{\Psi_{{#1},{#2},{#3}}}}     % for e^{abc}F^b F^c
\newcommand\vvs[1]{{\vartheta_{#1}}}            % for \delta^ab

\newcommand\uus[6]{{\Upsilon_{\{{#1},{#2},{#3}\},\{{#4},{#5},{#6}\}}}}
\def\kp{{\kappa}}

\def\cmp#1{{\it Comm. Math. Phys.} {\bf #1}}

\def\pl#1{{\it Phys. Lett.} {\bf #1B}}

\def\prd#1{{\it Phys. Rev.} {\bf D#1}}

\def\np#1{{\it Nucl. Phys.} {\bf B#1}}

\def\jmath#1{{\it J. Math. Phys.} {\bf #1}}
\def\mpl#1{{\it Mod. Phys. Lett.}{\bf A#1}}

\def\ap#1{{\it Ann. Phys.} {\bf #1}}
\def\hepth#1{{\it hep-th/{#1}}}

\begin{document}
\begin{titlepage}
\titlepage
\rightline{TAUP-2274-95}
\rightline{hep-th/9507036}
\rightline{July 6, 1995}
\vskip 1cm
\centerline {{\Large \bf The ``Dual" Variables Of Yang-Mills Theory}}
\centerline {{\Large \bf And Local Gauge Invariant Variables.}}
\vskip 1cm

\centerline {O. Ganor and J. Sonnenschein}

\vskip 1cm

\begin{center}

\em  School of Physics and Astronomy\\

Beverly and Raymond Sackler \\

Faculty of Exact Sciences\\

Tel-Aviv University\\

Ramat Aviv Tel-Aviv, 69987, Israel

\end{center}

\vskip 1cm

\abstract{
After adding auxiliary fields and integrating out the original variables,
the Yang-Mills action can be expressed in terms of local gauge invariant
variables. This method reproduces the known solution of the two dimensional
$SU(N)$ theory. In more than two dimensions the action splits into a
topological part and a part proportional to $\as$.
We demonstrate the procedure for $SU(2)$ in three dimensions where
we reproduce a gravity-like theory.
We discuss the four dimensional case as well. We use a cubic expression
in the fields as a space-time metric to obtain a covariant Lagrangian.
We also show how the four-dimensional $SU(2)$ theory can be expressed
in terms of a local action with six degrees of freedom only.
}
\end{titlepage}

%%%%%%%%%%%%%%%%%%%%%%%%%%%%%%%%%%%%%%%%%%%%%%%%%%%%%%%%%%%%%%%%%%%%%%%%%%%%
%                      BEGINNING OF PAPER                                  %
%%%%%%%%%%%%%%%%%%%%%%%%%%%%%%%%%%%%%%%%%%%%%%%%%%%%%%%%%%%%%%%%%%%%%%%%%%%%
\newtheorem{Lemma}{Lemma}
\tableofcontents

\section{Introduction}
One of the fascinating properties of certain quantum field theories is the
existence of a strong-weak duality of the coupling constant\cite{OlMo}.
Recently, various novel dualities were discovered both in the context of
supersymmetric gauge theories\cite{SeiWit,Sei} and topological field theories
\cite{VafWit}
 as well as in string theories\cite{WitSTR}.
In certain theories, like the compactified boson in 2D or the abelian  gauge
theory in 4D, the duality transformations  were performed
 by adding auxiliary fields and integrating
out the original variables\cite{Busche,RocVer,SeiWit}.

In this paper we explore this procedure in non-Abelian
non-supersymmetric Yang-Mills theories.
It is well known that the latter do not
 possess a
strong-weak duality invariance. (In fact such a duality is usually
meaningless when  the coupling constant is running.)  However,
the exciting exact results derived recently\cite{SeiWit} taught us that the
 study  of the resulting action in terms of the auxiliary
fields maybe very fruitful.

After the integration over the original variables, the resulting action
can be expressed in terms of {\em local gauge invariant} variables.
Having a gauge invariant description could be important for the large
$N$ limit of $SU(N)$ Yang-Mills theory. The correlation functions
of proper gauge invariant variables vanish as $1/N^2$ and thus
those variables can be considered classical
(Obviously, not every gauge invariant variable has this property).
This is the essence of the original master-field idea\cite{WittMF}.
In the approach of Migdal and Makeenko\cite{MigdMF,MakMig} the gauge
invariant variables are the non-local Wilson loops. More recently, Gross and
Gopakumar\cite{GopGro} suggested a different approach where the master
field is local but not gauge invariant. This master field corresponded
to the gauge field but as a non-commutative random variable.

In the present paper we obtain actions with local gauge invariant variables
for the $SU(2)$ theory.

In two-dimensions we obtain, by the above procedure, the $SU(N)$
partition function on a torus. This corresponds to the well-known
result\cite{MigdYM}\cite{Rusako} which is given as a sum over representations
of $SU(N)$. In our case, each representation corresponds to a different
configuration of the auxiliary field (whose value becomes quantized).

In more than two dimensions the action that we obtain splits into
a topological field theory plus a term proportional to the coupling
constant $\as$. The topological field theory describes the
flat gauge configurations at $\as=0$. We demonstrate this by
reproducing Lunev's result \cite{Lunv3D}
%%%\footnote{We thank D.Z.Freedman for familiarizing us with Lunev's result.}
for the three dimensional $SU(2)$  theory. It is expressed as
a theory similar to 3D gravity (the topological part)
plus a non-covariant coupling proportional to $\as$.

The topological field theory of flat gauge connections
was introduced in \cite{Cobim} in relation to 2D topological gravity.
In  \cite{Cobi}
theories of flat gauge connections of different groups and for
$D>2$
where written down. The 2D cases were then
discussed in \cite{WitGAG}\cite{WitREV}. In \cite{WitREV} the exact
instanton expansion of the 2D partition function was obtained by expressing
the action as a topological perturbation (proportional to $\as$) to the
flat gauge connection theory. The topological theory of flat gauge
connections in 4D was recently discussed in\cite{MarcN4} in relation
to a twisting of the super-symmetric $N=4$ Yang-Mills theory.

Continuing to 4D in gauge invariant variables, we obtain an expression
for the $SU(2)$ action in terms of a (non-positive) metric $g_{\u\v}$
and a chiral spin-2 field. We then show that we can restrict to conformal
metrics and thus obtain a description of $SU(2)$ in 4D in terms of a
local action of six gauge invariant fields only.
%%%%%%%%%%%%%%%

Historically, our procedure was pioneered
by Halpern\cite{Halp4D,HalpSD,HalpF}.
 In particular, our starting point is Halpern's 1977
field strength formulation\cite{Halp4D}
 of Yang-Mills. Halpern also used this
formulation to obtain a gauge invariant formulation\cite{HalpSD}
 of the self-dual
sector of Yang-Mills, which is a prototype of our gauge invariant formulation
of the full theory. The field strength formulation was also
employed  by Fradkin and Tseytlin\cite{FraTse}
%%%%%%%%%%%%%%%%
%The procedure that we described has already been applied to Yang-Mills
%heories by Halpern
%%% \cite{Halp4D,Halp3D,HalpSD,HalpF}
%\cite{Halp4D,HalpSD,HalpF}
%and by Fradkin and Tseytlin\cite{FraTse}.
%A gauge invariant formulation of self-dual fields was discussed in
%\cite{HalpSD}. In \cite{HalpF} a dual formulation in terms of the
%field strengths is written down.
%%% We mention again the related results of
%%% Lunev\cite{Lunv3D}\cite{Lunv4D}\cite{LunvGR}.
In \cite{Lunv3D} the gravitational description of 3D $SU(2)$ was found.
In \cite{Lunv4D} a gravity-like theory of 3D $SU(2)$ was found,
but it seems different from our description. In a recent work
\cite{LunvGR} more relations between gravity and Yang-Mills theories
are discussed.

The Hamiltonian approach to the gauge invariant description of Yang-Mills
is very interesting \cite{DZFHJL}\cite{BauDZF}\cite{HaaJoh}.
In this approach the metric is spatial and the wave functional can
be factorized\cite{DZFHJL}.

The paper is organized as follows. In section (2) we describe the
general framework. In section (3) we rederive the 2D partition function
of $SU(N)$ on a torus. In section (4) we discuss $SU(2)$ in 3D and
rederive Lunev's result\cite{Lunv3D}.
Section (5) is devoted to $SU(2)$ in 4D. We show how $SU(2)$ can be
expressed with only six fields -- a chiral spin-2 field and a spin-0 field.
Appendix (A) describes the topology of
the auxiliary fields for $SU(N)$ in 2D.
Appendix (B) describes an algebraic prescription to rewrite the action
in gauge invariant variables (which we use in section(4)).
Appendices (C-D) describes the detailed calculations for the 4D case.

\section{Reformulation of the theory}
We start with the Euclidean partition function
 in a  D dimensional space with  a metric $g_{\u\v}$
\bear
\lefteqn{
\ZFUN{D}{G} =\int\DAu\DFuv\DGtuv
exp \{-\int \sqrt{g}\{\fvas F_\uva F^\uva
}\nn\\
&+& i\dGX{\uv,a} (F_\uv^a - \pu A_\nu^a + \pv A_\mu^a
- t^{abc}A_\mu^b A_\nu^c)\}  d^Dx \}
\label{ZFUNAFG}
\eear
where $F_\uv^a$ is treated as an independent field,  $\dGX{\uv,a}$ is
an anti-symmetric Lagrange multiplier and $\alpha_s$ is the strong coupling
constant.
%%% In 4D
%%% \be
%%% \dGX{\uv,a} =\inv{2\sqrt{g}}\epu{\u\v\s\t}\GG{a}{\s\t}
%%% \ee
The anti-symmetric structure constants are defined as usual
\be
[T^a,T^b]      = i t^{abc} T^c
\ee

We proceed for simplicity on a flat toroidal space-time.
Integrating over $\DFuv$ we obtain
\be
\ZFUN{D}{G} =\int\DAu\DGtuv e^{-\int d^Dx \{\as\dG{a}{\uv}\dG{a}{\uv}
+2i\pu\dG{a}{\uv} A_\nu^a - i\dG{a}{\uv} t^{abc}A_\mu^b A_\nu^c)\}}
\label{ZFUNAG}
\ee
This action is quadratic in $A_\mu^a$ and the quadratic matrix
\be
\WG{ab}{\u\v}  = t^{abc}\dG{c}{\u\v} \label{QMAT}
\ee
is local in the new field variable $\dG{c}{\u\v}$.
For $D>2$ and a {\em generic} field configuration $\dG{c}{\u\v}$, it is also
a non-singular matrix and we can define its inverse $\VG{ab}{\u\v}$:
\be
\VG{ab}{\u\v}\WG{bc}{\v\t} = \dlu{ac}\dl{\u\t}
\ee
Integrating $\DAu$ we obtain
\be
\ZFUN{D>2}{G} =\int\DGtuv \prod_x (\det\cG(x))^{-1/2}
e^{-\int d^Dx \{\as \dG{a}{\uv}\dG{a}{\uv}
+ i\pu\dG{a}{\u\s}\VG{ab}{\s\t}\pv\dG{b}{\t\v} \}}
\label{ACTG}
\ee
The term $\pu\dG{a}{\u\s}\VG{ab}{\s\t}\pv\dG{b}{\t\v}$ is invariant
only up to a total derivative.
After adding a total derivative we can write the Lagrangian as:
\bear
\Lag &=& \as\dG{a}{\uv}\dG{a}{\uv} +
2i\dG{a}{\u\v}\VG{ab}{\v\t}\px{\u}\WG{bc}{\t\s}
       \VG{cd}{\s\r}\px{\a}\dG{d}{\a\r}
\nn\\
&-&
2i\dG{a}{\u\v}\VG{ab}{\v\t}\px{\u}\px{\s}\dG{b}{\s\t}
-i\VG{cd}{\v\r}\px{\a}\dG{c}{\a\v}\px{\b}\dG{d}{\b\r}
\nn\\
&=& \as
\dG{a}{\uv}\dG{a}{\uv}
+i\VG{cd}{\v\r}\px{\a}\dG{c}{\a\v}\px{\b}\dG{d}{\b\r} \nn\\
&-& 2i\px{\u} [\dG{a}{\u\v}\VG{ab}{\v\t}\px{\s}\dG{b}{\s\t}]
\nn\\
&&\label{FFB}
\eear
The same expression was obtained in\cite{Halp4D,FraTse}.
To see that it is gauge invariant we note the following observation.
If the {\em independent} variable $\dG{a}{\u\v}$ were a field strength
of some gauge field $\bA{a}{\u}$:
\be
\dG{a}{\u\v}    = \px{\u}\bA{a}{\v} -\px{\v}\bA{a}{\u}
                    -t^{abc}\bA{b}{\u}\bA{c}{\v}
\ee
and if it satisfied an equation of motion
(in 4D we could also use the Bianchi identity
and let $\GG{a}{\u\v}$ be the original field strength, but we
want to keep the dimension $D$ general):
\be
D_\u\dG{a}{\u\v} \eqdf
\px{\u}\dG{a}{\u\v} -t^{abc}\bA{b}{\u}\dG{c}{\u\v} = 0
\ee
we could solve for $\bA{a}{\u}$ and write
\be
\bA{a}{\u} \eqdf \VG{ab}{\u\v}\px{\t}\dG{b}{\t\v}
\ee
Clearly $\bA{a}{\u}$ transforms as a gauge field when $\dG{a}{\u\v}$
transforms in the adjoint representation.
We can thus define
\be
\bG{a}{\u\v}    = \px{\u}\bA{a}{\v} -\px{\v}\bA{a}{\u}
                    - t^{abc} \bA{b}{\u} \bA{c}{\v}
\ee
which transforms like $\dG{a}{\u\v}$ and we can write the
Lagrangian as
\be
\Lag =   \as\dG{a}{\u\v}\dG{a}{\u\v} - i\bG{a}{\u\v}\dG{a}{\u\v}
\ee
which is gauge invariant.
The terms in (\ref{FFB}) that contain the inverse of $\cG$
must be accompanied with a prescription for the pole when $\cG$ becomes
singular. For this purpose we have to add the gauge-breaking term
$-\h A^a_\u A^a_\u$ to the action before integrating, and
then take $\eta\rightarrow 0$.
This has the effect of replacing $\cG^{-1}$ with
\be
(\cG+i\h I)^{-1} =
P\frac{\mbox{adj}\cG}{\det\cG} -i\pi\delta(\det\cG)(\mbox{adj}\cG)
\mbox{sgn}(\tr{\mbox{adj}\cG})
\label{POLEPR}
\ee
where $\mbox{adj}\cG \eqdf (\det\cG)\cG^{-1}$.
The addition to $\bA{a}{\u}$ is
\be
-i\pi \delta(\det\cG) (\mbox{adj}\cG)^{ac}_{\u\v}\,\mbox{sgn}
(\tr{\mbox{adj}\cG})\px{\a}\dG{a}{\a\u}
\ee
and it is seen that this addition does not change the gauge
transformation properties of $\bA{a}{\u}$.
When we plug it in (\ref{FFB}) we obtain
\bear
\Lag = \as\dG{a}{\uv}\dG{a}{\uv}
&+& i\VG{cd}{\v\r}\px{\a}\dG{c}{\a\v}\px{\b}\dG{d}{\b\r}
- 2i\px{\u} [\dG{a}{\u\v}\VG{ab}{\v\t}\px{\s}\dG{b}{\s\t}] \nn\\
&+&\pi\delta(\det(\cG))
\,\mbox{sgn}(\tr{\mbox{adj}\cG})
(\mbox{adj}\cG)^{cd}_{\v\r}\px{\a}\dG{c}{\a\v}\px{\b}\dG{d}{\b\r}
\nn\\
&&\label{FFBPP}
\eear
It is enough to add the pole prescription just in
the term $\VG{cd}{\v\r}\px{\a}\dG{c}{\a\v}\px{\b}\dG{d}{\b\r}$
and not in the total derivative.
It is easy to check that the addition
\be
\pi\delta(\det(\cG))
\,\mbox{sgn}(\tr{\mbox{adj}\cG})
(\mbox{adj}\cG)^{cd}_{\v\r}\px{\a}\dG{c}{\a\v}\px{\b}\dG{d}{\b\r}
\label{ADDPP}
\ee
Is gauge invariant by itself, because of the $\delta$-function.

When the determinant $\det(\cG)$ contains multiple roots (as will
be the case for $SU(2)$ in $D=3$ in section (4) where the determinant
is a third power of an algebraic expression, and for $SU(2)$ in $D=4$
in section (5)  where the determinant will be quadratic) we will have
to be more careful. We will deal with such a situation in chapter (4).

We have seen that the terms in (\ref{FFB}) are local invariants.
However it is an expression that contains derivatives, but doesn't
contain any gauge field. Thus it must be possible to write it
in terms of local {\em algebraic} invariants of $\dG{a}{\u\v}$.
We will demonstrate this in various cases in the next sections.

At $\as =0$ we obtain a topological field theory that describes
flat  gauge connections\cite{Cobim,WitREV}. The Lagrangian
$- i\bG{a}{\u\v}\dG{a}{\u\v}$ thus describes a topological field
theory. Furthermore, all the manipulations above have been done
in a flat space. It is easy to see that if the term
$-\fvas\int d^Dx \{\F{a}{\uv}\F{a}{\uv}\}$ were missing from (\ref{ZFUNAFG})
we keep general covariance, since we can write the Lagrange multiplier
$\dG{a}{\uv}$ term in terms of forms only:
\be
\int \tr{G \wedge (F - dA - [A,A])} d^Dx
\ee
where $G$ is now a $(D-2)$ form.
The $\DAu$ measure
depends on the space-time metric through its volume form.
Since the integral is quadratic
and we just substitute the solution of the equations of motion,
we do not destroy covariance in the topological part of the action
\be
\Lag_{\mbox{topo}} = - i\bG{a}{\u\v}\GG{a}{\u\v}
\label{ACTOPO}
\ee
Thus, after we re-write $- i\bG{a}{\u\v}\GG{a}{\u\v}$
in gauge invariant variables, we obtain a topological field theory that
has general covariance.
When $\as=0$, only the term $\prod_x (\det\cG(x))^{-1/2}$ in (\ref{ACTG})
will break general covariance and behave as a density.
We will indeed see in section (4) that for three dimensional SU(2)
we reproduce Lunev's result \cite{Lunv3D} and obtain (a ``close cousin" of)
three dimensional gravity.
Three dimensional gravity itself is known to be a
topological field theory \cite{WittCS}.
For four dimensional $SU(2)$ we obtain another topological field theory.
Those theories describe flat gauge connections.

We will start with a discussion of  the well-understood two dimensional
Yang-Mills theory in this formalism.

\section{Rederiving the 2D partition function for $SU(N)$ on a torus}
Two dimensional Yang-Mills theory can be solved using  several different
methods
see for instance
 \cite{MigdYM,Rusako,WitREV,GopGro}.
The partition function on a torus of area $A$, for a gauge group $G$ is
\be
\ZFUN{2D}{G} = \sum_R e^{-\hlf\as A C_2(R)}
\label{ZC}
\ee
where $R$ runs over all representations of the group.
It is important to mention that (\ref{ZC}) is obtained after
we sum over all $G$-fibre bundles in the functional integral.
For a torus, the fibre-bundles are characterized by $\pi_1(G)$.
Consider an $SU(N)$  group for $G$. The difference between $SU(N)/Z_N$ and
$SU(N)$ in
(\ref{ZC}) manifests itself in the functional integral in
whether we sum over all $N=\#\,\pi_1(SU(N)/Z_N)$ bundles or just
the trivial bundle, since both algebras are the same.
The representations $R$ of $SU(N)$ are labeled by sets
of increasing natural numbers
$0<l_1<l_2<\cdots<l_{N-1}$. The second Casimir is given by
\be
C_2(l_1,l_2,\dots,l_{N-1}) =
\hlf\sum_{i=1}^{N-1} l_i^2 -\inv{2N}(\sum_{i=1}^{N-1} l_i)^2
\label{CASIM}
\ee
%%% \CHECK

\subsection{The $SU(2)$ partition function on a torus}
Let us begin with $SU(2)$ for which
\bear
a &=& 1,2,3 \nn\\
t^{abc} &=& \epu{abc} \nn\\
\dG{a}{\uv} &=& \ep{\uv} g^a
\eear
Choosing $g_a$ locally at $x$  to be in the positive $a=3$ direction, i.e.
$g^1 = g^2 = 0$ and $g^3 = g$,
the matrix of the quadratic term in $A_\mu^a$ in (\ref{QMAT}) is then:
\be
\left(
\begin{array}{cccccc}
0   & 0  & 0   &  0 & -ig & 0 \cr
0   & 0  & 0   & ig &  0  & 0 \cr
0   & 0  & 0   &  0 &  0  & 0 \cr
0   & ig & 0   &  0 &  0  & 0 \cr
-ig & 0  & 0   &  0 &  0  & 0 \cr
0   & 0  & 0   &  0 &  0  & 0
\end{array}
\right)
\label{CFMAT}
\ee
It has two zero eigenvectors which force two delta functions
$\prod_{\v=1,2}\delta(\pu\dG{3}{\uv})$, after an integration over $A_\u^3$.
The remaining integral (over $A_\u^1,A_\u^2$) is Gaussian and gives
\be
g^{-2}e^{-{i\over g}\ep{\s\t}(\pu\dG{1}{\u\s}
\pv\dG{2}{\v\t}-\pu\dG{2}{\u\s}\pv\dG{1}{\v\t})\Delta^2x}
\ee
In the case that $g_a$ is in a general iso-direction we write
\be
g^a = g \gh^a,\qquad \sum_{a=1}^3\gh^a\gh^a=1,\qquad g\ge 0
\ee
and obtain
\be
\prod_\v\delta(\gh^a\pu\GG{a}{\uv}) = \prod_\u\delta(\pu g)
\ee
where we used $\gh^a\pu\gh^a = 0$.
The product of delta-functions means that $g$ is a constant field.
However, there is an infinite constant involved because the arguments
of the delta functions are related by $\epu{\uv}\pu\pv g = 0$.
Since this infinite constant is independent of $\as$ or $g$
we disregard it!
We decompose the measure
\be
\DGuv = \prod_{x,a} d\GG{a}{\uv}(x) = \prod_x (g(x)^2 dg(x)( d^2\gh(x))
\ee
and obtain
\be
\ZFUN{2D}{SU(2)} =\int dg \{\int\Dgh e^{-2\as A g^2
-i g\int d^2x \epuv\epabc\gh^a\pu\gh^b\pv\gh^c}\}
\label{ZFUNGGH}
\ee
where $A$ is the area of the torus.
Since $\gh$ is a unit iso-vector, we have the identity
\be
8\pi n = \int d^2x \epuv\epabc\gh^a\pu\gh^b\pv\gh^c
\ee
where $n$ is the integer topological number of the map
$$
\gh : \Sigma \rightarrow S^2
$$
from our torus to $S^2$. More precisely $n$ is the rank of the map
$$
\gh^* : H^2(S^2,{\bf Z}) \rightarrow H^2(\Sigma,{\bf Z})
$$
The functional integral $\Dgh$ decomposes into a sum over integer $n$.
For each $n$ we have a number $v_n$ which is the ``volume" of the
subspace of maps $\gh^a$ with topological number $n$.
The $v_n$-s are universal constant numbers, {\em independent of} $\as$.
We assume that they are all equal, and we rescale them to one.
We obtain
\be
\ZFUN{2D}{SU(2)} =\int_0^\infty dg \sum_{n=-\infty}^\infty
e^{-2\as A g^2 + 8 g n\pi i}
\ee
where again the equality sign is up to a factor that is independent of
$\as$.
Next we write
\bear
\ZFUN{2D}{SU(2)} &=& \hlf\int_0^\infty dg  \sum_{n=-\infty}^\infty
(e^{-2\as A g^2 + 8 g n\pi i} + e^{-2\as A g^2 - 8 g n\pi i})
\nn\\
&=&
\hlf\int_{-\infty}^\infty dg \sum_{n=-\infty}^\infty
e^{-2\as A g^2 + 8 g n\pi i}
\eear
We finally obtain
\be
\sqrt{{\pi\over {8\as A}}}
\sum_{n=-\infty}^\infty e^{-\frac{8\pi^2}{\as A} n^2}
\ee
which, after a Poisson resummation,
differs from (\ref{ZC}) by a constant $\hlf$.
This constant comes from the $g=0$ contribution to the integral.
If we had been more careful with the prescription to pass around
the pole $g=0$ we would have obtained the extra $\hlf$.
We will show this now.

\subsubsection{The problem near $g=0$}
 From the formula
\bear
\ZFUN{\mbox{wrong}}{SU(2)} &=&
2\int_0^\infty dg \sum_{n=-\infty}^\infty
e^{-2\as A g^2 + 8 g n\pi i}
\nn\\
&=&
\hlf \int_0^\infty
 e^{-2\as A g^2}\sum_{m=-\infty}^\infty \delta(g-\frac{m}{4}) dg
\nn\\
&=& \sum_{m=1}^{\infty} e^{-\frac{\as A m^2}{8}} + \hlf
\label{ZFUNWR}
\eear
Where as we should obtain
\be
\sum_{j=0,\hlf,\dots} e^{-\frac{\as A}{8} (2j+1)^2}
\ee
So (\ref{ZFUNWR}) differs from the correct answer just by the
contribution of $m=0$ or, what is the same, from the $\delta(g)$ that
appears in the sum over $m$.
We claim that this $\delta(g)$ appears because of a wrong treatment of
the $g\approx 0$ region.

At $g=0$ the matrix $\cG$ at (\ref{CFMAT}) becomes zero.
For a regularization we add $-\h A^a_\u A^a_\u$ to the
original Lagrangian. We thus have
\be
\Lag =
2\as g^a g^a - 2i\ep{\u\v}\px{\u}g^a A^a_\v
-(ig^a\epu{abc}\ep{\u\v} +\h\dlu{bc}\dl{\u\v}) A^b_\u A^c_\v
\ee
The inverse of the quadratic matrix
\be
\bM^{bc}_{\u\v} = ig^a\epu{abc}\ep{\u\v} +\h\dlu{bc}\dl{\u\v}
\ee
is
\be
(\bM^{-1})^{ab}_{\u\v} =
\inv{\h^2 + g^2}(\h \dlu{ab}\dl{\u\v}
      -ig^c\epu{abc}\ep{\u\v} +\h^{-1}g^a g^b \dl{\u\v})
\ee
Integrating out $A^a_\u$ we obtain
\be
+\frac{\h}{\h^2+g^2}|\px{\u}g^a|^2
+\inv{4\h(\h^2+g^2)} |\px{\u}(g^2)|^2
-\frac{i}{\h^2+g^2}\ep{\u\v}\epu{abc}g^a\px{\u}g^b\px{\v}g^c
\ee
For $g\gg\h$ the second term produces the delta function
that forces $g^2 =\const$, while the first term can be ignored
and the last term produces the topological invariant.
We see that for $g\gg\h$ indeed all topological sectors appear
with the same weight (since the coefficient of $|\px{\u}g^a|^2$
in the first term is negligible).
Thus for $g\neq 0$ we indeed get $\sum_{m\neq 0} \delta(g-\frac{m}{4})$.
However as $g\approx\h$  the first term $|\px{\u}g^a|^2$ damps the
fluctuations of $\gh^a$.
The result of this is that the sum $\sum_n e^{8\pi i g n}$ is finite,
and we do not get the $\delta(g)$ term.

\subsection{Generalization to $SU(N)$}
Let us choose a Cartan subalgebra of the Lie algebra and denote
its generators by $T^i$ with $i=1,\dots,N-1$. The rest of the
roots will be denoted by $T^\ra$.
As in the $SU(2)$ case, where we parameterized
$\hlf\ep{\u\v}\dG{a}{\u\v}$ by $g$ and $\gh^a$ we now
parameterize $\ep{\u\v}\dG{a}{\u\v}$ by
\be
\dG{a}{\u\v} = \hlf\ep{\u\v} \sum_{i=1}^{N-1} H^i (\gh T^i \gh^{-1})^a
\ee
where $\sum_i H^i T^i$ is an element of the Cartan subalgebra
which is conjugate to $\dG{a}{\u\v}$. For generic $\dG{a}{\u\v}$,
the $H^i$-s are unique up to the Weyl group $S_N$.
$\gh$ is the (generically) unique element from the coset
\be
\gh \in SU(N)/U(1)^{N-1}
\ee
where $U(1)^{N-1}$ represents the Cartan torus in $SU(N)$ that corresponds
to our choice of Cartan subalgebra.
$\gh$ is represented by $\gh\in SU(N)$ such that $\gh\equiv \gh h$
for $h\in U(1)^{N-1}$.
The matrix $\cG(x)$ has now $2(N-1)$ zero modes, two for each
Cartan direction $T^i$. The integration over $A^i_\u$ in the zero
mode direction will, as before, produce a constancy condition:
\be
\pu H^i = 0,\qquad i=1,\dots,N-1
\ee
Choosing locally at x, $\gh(x)=1$ we find that
the remaining terms decouple for each positive root $\ra$:
\be
iA^\ra_\v \ep{\u\v}\pv G^{-\ra} +
iA^{-\ra}_\v \ep{\u\v}\pv G^{\ra}
-i\ep{\u\v}(\sum_i \ra(i) H^i)A^\ra_\v A^{-\ra}_\u
\ee
where
\be
G^a \eqdf \hlf\ep{\u\v}\dG{a}{\u\v}
\ee
After the integration we have in the exponent
\bear
\lefteqn{
\sum_\ra \inv{\sum_i \ra(i) H^i}
\ep{\u\v} \pv G^{\ra}  \pu G^{-\ra}
} \nn\\
&=&
\sum_\ra \inv{\sum_i \ra(i) H^i}
\ep{\u\v}
\sum_{i,j}H^i H^j \pv\tr{\gh T^i\gh^{-1}T^\ra}
\pu\tr{\gh T^i\gh^{-1}T^{-\ra}}
\nn\\
&& \label{EXPN}
\eear
and in front of the exponent we have
\be
\inv{\prod_\ra (\sum_i \ra(i) H^i)^2}
\ee
which cancels with the Jacobian for passing from
$\prod_a dG^a$ to $\prod_i dH^i d^{(N^2-N)}\gh$.
In fact, it is the well known square of the Van-Der-Monde of $H^i$
in an appropriate basis for the Cartan subalgebra.
%%% \CHECK

At $x$, since $\gh(x)=1$ we have
\bear
\sum_i H^i \pu\tr{\gh T^i\gh^{-1}T^\ra}
&=& -\sum_i H^i \tr{T^i [T^\ra,\gh^{-1}\pu\gh] }
\nn\\
=  \sum_i H^i \tr{[T^i, T^\ra] \gh^{-1}\pu\gh }
&=&  (\sum_i \ra(i)H^i) \tr{T^\ra \gh^{-1}\pu\gh }  \nn
\eear
The exponent (\ref{EXPN}) reads, at $x$,
\bear
\sum_\ra (\sum_i \ra(i)H^i) \ep{\u\v}
\tr{T^{-\ra}\gh^{-1}\pv\gh}\tr{T^\ra\gh^{-1}\pu\gh}
\eear
Since
\be
\gh^{-1}\pu\gh = \sum_i T^i \tr{T^i\gh^{-1}\pu\gh}
+\sum_\ra T^{-\ra} \tr{T^\ra\gh^{-1}\pu\gh}
+\sum_{\ra} T^\ra \tr{T^{-\ra}\gh^{-1}\pu\gh}
\ee
we can write the exponent as
\be
i\ep{\u\v}\sum_i H^i\tr{T^i [\gh^{-1}\pu\gh,\gh^{-1}\pv\gh]}
\ee
But the expressions
\be
\inv{2\pi}\int \ep{\u\v}\tr{T^i [\gh^{-1}\pu\gh,\gh^{-1}\pv\gh]}d^2x
\label{CHRNI}
\ee
%%% \CHECK
are integers. In fact (see Appendix (A)) they are the pullbacks by
$\gh$ of a basis of the second integer cohomology group
$H^2(SU(N)/U(1)^{N-1},Z)$ back to the torus.
So we write
\be
n_i = \inv{2\pi}\int \ep{\u\v}\tr{T^i [\gh^{-1}\pu\gh,\gh^{-1}\pv\gh]}d^2x
\ee
and $n_1,\dots,n_{N-1}$ characterize the topological sector of the
map $\gh$.
We are left with:
\be
\sum_{\{n_i\}}
\int\inv{N!}\prod_i dH^i e^{-2\as\sum_{i,j} H^i H^j \tr{T^i T^j}
-8\pi i\sum_i H^i n_i}
\ee
Using   (see Appendix (A))
\be
\tr{T^i T^j} = \dlu{ij}-\inv{N}
\ee
and substituting
\be
\sum_{n_i} e^{-8\pi i H^i n_i} = \inv{4}\sum_{l_i} \delta(H^i - \inv{4}l_i),
\ee
we obtain
\be
\inv{4^{N-1}N!}\sum_{\{l_i\}}
e^{-\inv{2}\as (\sum_{i=1}^{N-1} l_i^2 -\inv{N}(\sum_{i=1}^{N-1} l_i)^2)}
\label{SUMSUN}
\ee
where the $N!$ is the size of the Weyl group.
%%% \CHECK
Again the remaining terms in (\ref{ZC}-\ref{CASIM}) come from carefully
taking care of the points where the matrix $\cG$ is more singular than
usual. These are points where $\sum_i\alpha(i) H^i = 0$ for some root
$\alpha$. In our case, these are points where $H^i=H^j$ for some $i\ne j$,
or $H^i=0$ for some $i$.
Those points must be excluded from the $\delta$-function
as in the $SU(2)$ case.
Defining
\be
D(m_1,\dots,m_N) \eqdf \inv{N}\sum_{i<j} (m_i-m_j)^2
\ee
We can write
\be
\sum_{i=1}^{N-1} l_i^2 -\inv{N}(\sum_{i=1}^{N-1} l_i)^2
=D(0,l_1,\dots,l_{N-1})
\ee
Noting that $D(m_1,\dots,m_N)$ is invariant under permutations and
that for any integer $m$:
\be
D(m_1,\dots,m_N) = D(m_1-m,\dots,m_N-m)
\ee
We see that we can write (\ref{ZC}-\ref{CASIM}) as
\be
\inv{N!}\sum_{\begin{array}{c}
              l_1,\dots,l_{N-1} \\ l_i\neq 0, l_i\neq l_j
              \end{array}}
              e^{-\hlf\as A D(0,l_1,\dots,l_N)}
\ee
Which is the same as (\ref{SUMSUN}) with the above restrictions
on $H^i$-s.

\section{Three dimensional $SU(2)$}
Proceeding to three dimensions, we wish to rewrite the YM  Lagrangian
 in terms of only gauge invariant variables.
The following analysis resembles the case of self-dual Yang-Mills
in 4D given in \cite{HalpSD}.
%%% \subsection{Passing to gauge invariant variables}

We define the `` magnetic'' field $B^a_i$ by
\be
\dG{a}{ij} = \ep{ijk}B^a_k
\ee
The final variables will be bilinear in the field strength.
We further define the symmetric semi-positive matrix
\be
T_{ij} \eqdf B^a_i B^a_j
\label{TIJ}
\ee
We mention in advance a puzzle that might arise about our plan.
It is known\cite{WuYang,DZFKhu}
%%%\footnote{We wish to thank D.Z.Freedman for pointing our attention
%%%to the Wu-Yang ambiguity.}
that in three dimensions the invariant bilinears
$T_{ij}$ are not enough to completely describe a gauge configuration
$A^a_i$ that yields $B^a_i$.
Namely, there can be two different configurations of $A^a_i$
with identical $T_{ij}$ but different $B^a_i (D_j B_k)^a$
where $D_j$ is the covariant derivative (although this is not the
generic situation) \cite{DZFKhu}.
In our case, however, it is important to remember that the $T_{ij}$-s
are not bilinear in the original field strength but in the
{\em auxiliary} fields, thus there is no conflict.

\subsection{The action}
In general the inverse matrix $\cG^{-1}$ can be expressed as
a rational function in $\dG{a}{\u\v}$ with a denominator that
is $\det(\cG)$ of degree $D(N^2-1)$.
For three-dimensional $SU(2)$ we can reduce the degree from $9$
to $3$. This is related to the fact that $\det(\cG)$ is a third
power of a cubic polynomial.
Defining
\be
\DD \eqdf {1\over 6}\epu{abc}\ep{ijk} B^a_i B^b_j B^c_k
\ee
we have
\be
\DD^2 = \det(T_{ij})
\ee
we get
\be
\VG{ab}{ij} = \inv{2\DD}(B^a_i B^b_j - 2 B^a_j B^b_i)
\ee
Plugging this in (\ref{FFB}) we obtain
%%% \bear
%%% \ep{ijp}B^a_p\bG{a}{ij} &=&
%%% \inv{\DD}\ep{klm}\ep{pji}T_{pk}\px{i}\px{m}T_{jl}
%%% \nn\\
%%% +\inv{4\DD}D_{ij}T_{st}(\ep{slp}\ep{tmk}-\ep{spk}\ep{tlm})
%%% \px{p}T_{ki}\px{m}T_{lj}
%%% &+&\inv{\DD}T_{kp}\ep{klm}\ep{pij}\px{m}T_{lj}\px{i}\log\DD
%%% \nn\\
%%% &+&\frac{3}{4\DD}\ep{klm}\ep{pij}\px{p}T_{ki}\px{m}T_{lj}
%%% \nn\\
%%% &&
%%% \eear
\bear
\ep{ijp}B^a_p\bG{a}{ij} &=&
\inv{\DD}\ep{klm}\ep{pji}T_{pk}\px{i}\px{m}T_{jl}
+\inv{\DD}T_{kp}\ep{klm}\ep{pij}\px{m}T_{lj}\px{i}\log\DD
\nn\\
&+& \inv{4\DD}D_{ij}T_{st}(\ep{slp}\ep{tmk}-\ep{spk}\ep{tlm})
\px{p}T_{ki}\px{m}T_{lj}
+\frac{3}{4\DD}\ep{klm}\ep{pij}\px{p}T_{ki}\px{m}T_{lj}
\nn\\
&&
\eear

where
\bear
%%\DD    &=& \inv{6}\epu{abc}\ep{ijk}B^a_i B^b_j B^c_k   \nn\\
%%T_{ij} &=& B^a_i B^a_j                                 \nn\\
D_{ij} &=& \inv{2\DD^2}\ep{jst}\ep{ilm}T_{sl}T_{mt}    \nn\\
T_{ij}D_{jk} &=& \dl{ik}                               \nn
\eear
After integration by parts  the topological part of the action takes the form
\bear
S &=&
i\int
\{
\inv{4\DD}D_{ij}T_{st}(\ep{slp}\ep{tmk}-\ep{spk}\ep{tlm})
\px{p}T_{ki}\px{m}T_{lj}
\nn\\
&-&\inv{4\DD}\ep{klm}\ep{pij}\px{p}T_{ki}\px{m}T_{lj}
\}
\eear
This reproduces the result of Lunev\cite{Lunv3D} since it is
easy to check that if we take $g_{ij}=T_{ij}$ to be the metric, then we
can write
\be
\Lag = i\sqrt{g} R
\ee
where $R$ is the curvature built out of the metric.
We see that we get a topological theory (3D gravity) as expected.

There are still two more things to be taken care of:
The first has to do with the fact that $\DD$ can have either $\pm$ sign,
where as $\sqrt{g}$ is defined to be positive.
The second and related problem, is the prescription of passing
around zeroes of $\DD$.
%%% We will return to those problems shortly,
%%% but first we discuss the Jacobian.

\subsection{Going round $\DD=0$}
It remains to calculate the analog of (\ref{ADDPP}). Since
$\det(\cG)$ is a cubic power, (\ref{ADDPP}) has to be modified.
We can write
\bear
\VG{ab}{ij} &=& \inv{\DD} K^{ab}_{ij} \nn\\
K^{ab}_{ij} &\eqdf&  \hlf(B^a_i B^b_j - 2 B^a_j B^b_i) \nn
\eear
Now we can write the addition due to the pole prescription as
\be
\delta\Lag_{PP} = \pi \delta(\DD)
\,\mbox{sgn}(\tr{\mbox{adj}K}) K^{ab}_{ij}
\ep{ipk}\px{p}B^a_k \ep{jql}\px{q}B^b_l
\ee
We can put
\be
\mbox{sgn}(\tr{\mbox{adj}K}) = -sgn( B^a_i B^a_i) = -1
\ee
and obtain
\be
\delta\Lag_{PP} = -\pi \delta(\DD)
K^{ab}_{ij} \ep{ipk}\px{p}B^a_k \ep{jql}\px{q}B^b_l
\label{PP1}
\ee
which is invariant at points $x$ where $\DD(x)=0$.
At such a point there is a direction $\hat{n}_i$ such that
\be
B^a_i \hat{n}_i = 0,\qquad a=1,2,3
\ee
We can write the invariant projection matrix on this direction as
\be
P_{ij}\eqdf\hat{n}_i\hat{n}_j
=
\frac{T_{ik} T_{kj} - T_{kk} T_{ij}}{\hlf (T_{kk}T_{ll} - T_{kl}T_{lk})}
+\dl{ij}
\ee
which can easily be checked by diagonalizing $T_{ij}$.
Next, choose for simplicity $\hat{n}$ in the $\hat{x}_3$ direction,
i.e. $\hat{n}_i = \dl{i3}$.
We denote by $\hi,\hj,\hk,\dots$ indices that run only on $1,2$.
We further denote
\be
\ep{\hi\hj}\eqdf\ep{\hi\hj 3}
\ee
and use
\bear
K^{ab}_{\hi\hj}\ep{\hj\hl}\px{3}B^b_\hl
&=& -\hlf \ep{\hj\hl}B^a_\hj\px{3}T_{\hl\hi}
\qquad\mbox{at $x=0$} \nn\\
\px{i}T_{j3} &=& B^a_j\px{i}B^a_3 \qquad\mbox{at $x=0$} \nn
\eear
We find
\bear
K^{ab}_{ij} \ep{ipk}\px{p}B^a_k \ep{jql}\px{q}B^b_l
&=& -\inv{4}\ep{\hj\hl}\ep{\hi\hk}\px{3}T_{\hj\hk}\px{3}T_{\hl\hi} \nn\\
 +  \ep{\hj\hl}\ep{\hi\hk}\px{\hk}T_{\hj 3}\px{3}T_{\hl\hi}
&+& \hlf \ep{\hj\hl}\ep{\hi\hk}\px{\hk}T_{\hi 3}\px{\hl}T_{\hj 3}
 -  \ep{\hj\hl}\ep{\hi\hk}\px{\hk}T_{\hj 3}\px{\hl}T_{\hi 3} \nn
\eear
Using the invariant projection operator we can write this as
\bear
\delta\Lag_{PP} &=& -\pi \delta(\DD)
P_{pq}P_{st}\ep{jlp}\ep{ikq}(\px{k}T_{js}\px{t}T_{li} \nn\\
&-&\inv{4}\px{s}T_{jk}\px{t}T_{li}
+\hlf \px{k}T_{is}\px{l}T_{jt}- \px{k}T_{js}\px{l}T_{it}) \nn\\
P_{ij}\eqdf\hat{n}_i\hat{n}_j
&=&
\frac{T_{ik} T_{kj} - T_{kk} T_{ij}}{\hlf (T_{kk}T_{ll} - T_{kl}T_{lk})}
+\dl{ij} \nn\\
&&\label{POLED3}
\eear

\subsection{The full functional integral}
%%% \subsection{The Jacobian}
The local change of variables from $B^a_i$ to $T_{ij}$
is of course accompanied by a Jacobian which joins with the
factor
\be
\inv{\sqrt{\det(\cG)}} = \DD^{-3/2}
\ee
%%% We will discuss with the sign $\pm$ of the square in the next subsection.
The Jacobian of the transformation is
\be
J(T)\eqdf \int \prod_{i\leq j} \delta(T_{ij}-B^a_i B^a_j) \prod dB^a_i
=\const\times\det(T)^{-1/2}=\const\times|\DD|^{-1}
\ee
%%% It can be obtained in a standard way,
%%% but in order to see this quickly we can calculate
%%% \be
%%% \int e^{-\sum M_{ij} T_{ij}}J(T)\prod_{i\leq j} dT_{ij}
%%% =\const\times(\det M)^{-3/2}
%%% \ee
%%% where $M$ is a positive symmetric matrix.
%%% On the other hand, we have for any matrix $\Lambda_{ij}$
%%% \be
%%% \int e^{-\tr{\Lambda^t T \Lambda}}\det(T)^k\prod_{i\leq j} dT_{ij}
%%% = \const \times (\det (\Lambda\Lambda^t))^{-2-k}
%%% \ee
%%% by making the change of variables $T\rightarrow \Lambda^t T\Lambda$
%%% whose Jacobian is $(\det\Lambda)^4$
%%% %% (by checking for infinitesimal $\Lambda$).
%%% So
%%% \be
%%% J(T) = \const\times\det(T)^{-1/2}=\const\times|\DD|^{-1}
%%% \ee

%%% \subsection{The full functional integral}
The functional integral now looks like
\be
\sum_{sgn(\DD(x))}
\int\cD{T_{ij}} \prod_x ({1\over{\sqrt{\DD^3}}})
exp\{-\int({{i S(T_{ij})}\over\DD} -\as\tr{T})\}
\label{SIGNDD}
\ee
where the generally covariant integration measure is
\be
\cD{T_{ij}} \eqdf \prod_x(\inv{\sqrt{T}}\prod_{i\leq j} dT_{ij})
              =\prod_x(|\DD|^{-1}\prod_{i\leq j} dT_{ij})
\ee
There are several problems with this action because of
the sum over the sign of each $\DD$.
The entire space is divided into regions separated by surfaces
where $\DD=0$. In each region, we have to sum over the {\em global}
sign of $\DD$ in that region. In the vicinity of the surfaces of
zero $\DD$ we have to put in the addition (\ref{POLED3}).
The problems arise because of two reasons. First, when we choose the
negative sign for $\DD$,
the term ${1\over{\sqrt{\DD^3}}}$ contains an extra $i$. Since we
have to multiply those terms for every $x$, there is a phase
ambiguity. Furthermore, at the boundaries $\DD=0$ and
${1\over{\sqrt{\DD^3}}}$ gives an infinite contribution.

We will now propose a way out, though we do not know if it is really
well defined. The problems really started because we changed the
order of the integration in (\ref{ZFUNAFG}), which is not absolutely
converging.  The pole prescription (\ref{POLEPR}) makes the integral
absolutely converging but is not enough to settle the phase ambiguity
in (\ref{SIGNDD}). We propose to put the problematic term
${1\over{\sqrt{\DD^3}}}$ back into the exponential where it came from,
by adding an {\em invariant} field $\phi$ and write
\be
\sum_{sgn(\DD(x))} \int\cD\phi\int\cD{T_{ij}}
exp\{-\int({{i S(T_{ij})}\over\DD} -\pi\delta(\DD)P(T_{ij})-\as\tr{T}
-i\DD^3\phi^2) \}
\ee
where we added the pole prescription term $\pi\delta(\DD)P(T_{ij})$
from (\ref{POLED3}). The integration over $\phi$ should be performed
at the end, and can be done with a regulator $-\h\phi^2$.

\section{Four dimensional $SU(2)$}
In four dimensions, our invariant variables will again be
bilinear expressions in the field strength. However, as there are
$D(D-1)/2=6$ field strengths, there are $21$ bilinear variables.
Under $O(4)$ they decompose into real representations as
\be
21 = \rep{1}\oplus\rep{1}\oplus\rep{9}\oplus\rep{10}
\ee
(the two $\rep{1}$-s are $\GG{a}{\u\v}\GG{a}{\u\v}$
and $\dG{a}{\u\v}\GG{a}{\u\v}$, the $\rep{9}$ is the traceless symmetric
tensor $\GG{a}{\u\t}\GG{a}{\t\v} - \mbox{(trace)}$, and
$\rep{10}=\rep{5}\oplus\rep{5}^*$ where ($\rep{5}^*$)~$\rep{5}$
is the product of two (anti-~)self-dual parts (minus trace).
The number of degrees of freedom minus gauge degrees is
$3*6-3=15$ so we have to use an overcomplete set of variables if we
wish to use only bilinear variables and keep $O(4)$ invariance.

Although the representation $\rep{5}$ describes self-dual tensors,
there is a way to use it without the anti-self-dual parts.
We will see shortly that a combination that is cubic
in the field strength describes a (non-positive) metric which makes
the field strengths $\GG{a}{\u\v}$ self-dual!

%%% \subsection{Notation}
%%% For convenience, we denote by a single index (e.g. $i_1$)
%%% a pair of indices with respect to which a quantity is
%%% anti-symmetric.
%%% Thus if
%%% \be
%%% i_1 = \{\u\v\},\qquad
%%% i_2 = \{\v\u\},\qquad
%%% i_3 = \{\s\t\}
%%% \ee
%%% then in all the quantities that we will use, if
%%% $i_1$ appears as an index an odd number of times in the quantity,
%%% then replacing it by $i_2$ changes the sign of the quantity.
%%% For example, the field strength $\F{a}{i_1}=-\F{a}{i_2}$ depends on
%%% one index. We also define
%%% \bear
%%% \dlt{i_1}{i_3} &\eqdf&
%%% \hlf (\dlt{\u}{\s} \dlt{\v}{\t}-\dlt{\u}{\t} \dlt{\v}{\s})
%%% =\hlf \dlt{[\u}{\s} \dlt{\v]}{\t} \nn\\
%%% \ep{i_1 i_3} &\eqdf& \hlf\ep{\u\v\s\t} \nn
%%% \eear
%%% where $[\cdots]$ means complete anti-symmetrization of anything
%%% in between.
%%% The dual field strength is
%%% \be
%%% \GG{a}{i_1}=\GG{a}{\u\v} \eqdf \hlf\ep{\u\v\s\t}\dG{a}{\s\t}
%%% =\ep{i_1 i_3}\dG{a}{i_3}
%%% \ee
%%% We define the gauge invariants:
%%% \bear
%%% T_{12} &=& \dG{a}{1}\dG{a}{2}                      \label{TINVAR}\\
%%% K_{123} &=& \epu{def}\dG{d}{1}\dG{e}{2}\dG{f}{3}   \label{KINVAR}
%%% \eear
%%% $T$ is symmetric in its two indices and contains $21$ fields
%%% and $K$ in anti-symmetric and contains $20$ fields.
%%% in its indices.

\subsection{Algebraic identities}
Our aim is again to write (\ref{FFB}) in terms of gauge invariant variables.
We start with some interesting algebraic properties of our $SU(2)$ variables.
The algebraic facts in this subsection, can all be checked with
a tedious but straightforward calculation.

Recalling that
\be
\GG{a}{\u\v} = \hlf\ep{\u\v\s\t}\dG{a}{\s\t}
\ee
The equation for the inverse is:
\bear
\VG{ab}{\u\v} &=&
\inv{\DD}[\inv{8}(\GG{a}{\a\b}\dG{b}{\g\d}-\GG{b}{\a\b}\dG{a}{\g\d})
(\epu{def}\GG{d}{\u\v}\dG{e}{\a\b}\GG{f}{\g\d})
\nn\\
&-&\inv{6}\GG{a}{\g\d}\dG{b}{\g\d}
(\epu{def}\GG{d}{\u\a}\dG{e}{\a\b}\GG{f}{\b\v})]   \label{INVF} \\
\DD &=&
\inv{48} (\epu{abc}\GG{a}{\a\b}\GG{b}{\g\d}\GG{c}{\s\t})
(\epu{def}\dG{d}{\a\b}\dG{e}{\g\d}\dG{f}{\s\t})
\label{DEFDD}
\eear
The determinant is
\be
\det(\cG) = \frac{\DD^2}{4}
\ee
Those equations were also found in \cite{DesTei,Halp4D}.

We further define a symmetric tensor
\bear
g_{\u\v} &=& \inv{3\DD^{1/3}}
\epu{def}\GG{d}{\u\a}\dG{e}{\a\b}\GG{f}{\b\v} \label{GINVAR}\\
g^{\u\v} &=& \frac{2}{3\DD^{2/3}}
\epu{def}\dG{d}{\u\a}\GG{e}{\a\b}\dG{f}{\b\v} \label{GUINVA}
\eear
As the notation suggests $g_{\u\v}$ is a symmetric
tensor (10 components) and $g^{\u\v}$ is its inverse.
We have
\be
\det_{4\times 4}(g_{\u\v}) = \inv{4}\DD^{2/3}
\ee
The $\DD$ scalings in (\ref{GINVAR}-\ref{GUINVA}) have been chosen
so that $g_{\u\v}$ will be a covariant tensor.
The metric (\ref{GINVAR}) has the important property that it makes
the field strengths {\em self-dual}.
\be
\inv{2\sqrt{g}}\epu{\a\b\t\s}g_{\u\a}g_{\v\b}\GG{a}{\t\s} = \GG{a}{\u\v}
\label{ANTISDG}
\ee
Actually, the distinction between self-dual and anti-self-dual here
is just what sign  we take in $\sqrt{g}=\pm\hlf\DD^{1/3}$.
Since the metric $g_{\u\v}$ is not necessarily positive definite,
we can choose either sign, and we need to sum over a {\em global}
sign, just like the summation over signs in the 3D case in (\ref{SIGNDD}).

\subsection{Gauge invariant variables}
Up to now we have not paid much attention to the distinction between
covariant and contra-variant tensors, because we were working
in flat Euclidean space-time. Now that we have chosen the metric
(\ref{GINVAR}) we will make this distinction.
The original field $\GG{a}{\u\v}$ is by definition covariant.
We now define
\bear
\Fd{a}{\u\v} &\eqdf& \GG{a}{\u\v}
        =\inv{\sqrt{g}} g_{\u\a} g_{\v\b} \widetilde{G}^{\a\b,a} \nn\\
\Fu{a}{\u\v} &\eqdf& g^{\u\a}g^{\v\b}\GG{a}{\a\b}
        =\inv{\sqrt{g}} \widetilde{G}^{\a\b,a} \nn
\eear
{}From now on, raising and lowering of indices are with respect to
(\ref{GINVAR}-\ref{GUINVA}).

We can write (\ref{INVF}) as:
\be
\VG{ab}{\u\v} = -\hlf g^{\a\b}\Fd{a}{\u\a}\Fd{b}{\b\v}
                -\inv{4}\Fu{a}{\a\b}\Fd{b}{\a\b}\, g_{\u\v}
\ee

The $10$ fields of the metric $g_{\u\v}$ plus the $5$ fields
of the traceless, $g_{\u\v}$-self-dual tensor
\bear
W_{\u\v\s\t} &=&
\Fd{a}{\u\v}\Fd{a}{\s\t} -
\inv{24\sqrt{g}}\epu{\a\b\g\d}\Fd{a}{\a\b}\Fd{a}{\g\d}
(g_{\u\s}g_{\v\t}-g_{\u\t}g_{\v\s} +\sqrt{g} \ep{\u\v\s\t}) \nn\\
&& \label{WDEF}
\eear
%%% \CHECK
form $15$ independent variables, which is exactly the number
of degrees of freedom that are left after eliminating the gauge
degrees.
We note, however, that the metric (\ref{GINVAR}) is not necessarily
positive definite.

The topological Lagrangian
(\ref{FFB}) can be written in terms of the ``metric"
and the field $W_{\u\v\s\t}$.
The resulting expression is rather cumbersome, and is described
in appendix (C). We define
\be
\Phi \eqdf \inv{24\sqrt{g}} \epu{\u\v\s\t}\GG{a}{\u\v}\GG{a}{\s\t}
\label{PHIDEF}
\ee
$\Phi$ is a scalar field which, given that metric, can be written
in terms of the chiral spin-2 field $W$. We show this in Appendix (D).

The non-covariant $\as$-dependent term is
\bear
&& \as\GG{a}{\uv}\GG{a}{\uv}
= \as \dlu{\u\s}\dlu{\v\t}\dG{a}{\u\v}\dG{a}{\s\t}
\nn\\
&=&
\as \dlu{\u\s}\dlu{\v\t} \{W_{\u\v\s\t} +
(g_{\u\s}g_{\v\t}-g_{\u\t}g_{\v\s})\Phi\}
\eear
where we used (\ref{WDEF}) and (\ref{PHIDEF}).

\subsection{Reduction to conformal metrics}
So far we have a description with $10+5$ fields.
However, we can in fact, restrict to conformal metrics.
What happens if in (\ref{ZFUNAFG}) we restrict $\dG{a}{\u\v}$ to be
self-dual? This corresponds to an original Yang-Mills action
\be
\inv{4\as}\int\F{a}{\u\v}\F{a}{\u\v}d^4x
-\inv{4\as}\int\F{a}{\u\v}\dF{a}{\u\v}d^4x
\ee
Note that $\theta$ angle is imaginary.
When we restrict to a trivial $SU(2)$-bundle, the instanton number is zero
and we may drop the $\theta$-term. For non-trivial bundles, we have to
be more careful in the $\DAu$ integration in (\ref{ZFUNAFG}) that
produces (\ref{FFB}) .We will   elaborate on those matters in a later
work\cite{INPREP}.
Note that when the $\GG{a}{\u\v}$-s are self-dual
$\sqrt{g}\ne 0$ everywhere, the instanton number must be
trivial since  the three fields $\GG{a}{14},\GG{a}{24},\GG{a}{34}$
are linearly independent and establish a {\em frame} of the associated
rank-3 vector bundle.

When $\dG{a}{\u\v}$ is self dual, the metric (\ref{GINVAR}) is conformal:
\be
g_{\u\v} = \psi \dl{\u\v}
\ee
(This fact was already pointed out in \cite{Halp4D}).
The action can thus be expressed in terms of the six fields: the
spin-2 self-dual $W_{\u\v\s\t}$ (5 fields) and the spin-0 $\psi$.
The resulting action can be derived from the formula in appendix (C).
This will be  described  in more detail in a future publication\cite{INPREP}.

\section{Discussion}
Extending early work by Halpern\cite{Halp4D-HalpFF}
we have explored the description of Yang-Mills theory
in terms of the auxiliary ``dual" variables.
We have seen that in two dimensions this formalism reproduces the
known solution\cite{MigdYM,Rusako} of the torus partition function.
The conjugacy class of the dual field becomes constant (over space)
and quantized. The sum over representations
of $SU(N)$ in \cite{MigdYM,Rusako} corresponds to a sum over the quantized
values of the conjugacy class of the auxiliary field.
This may be compared to an instanton expansion\cite{WitREV} which
gives the Poisson resummed partition function.

In the  three dimensional $SU(2)$
gauge theory,  we showed that the
action can be expressed as a sum of 3D gravity\cite{Lunv3D} plus a
non-covariant coupling. We have seen that there is an extra non-covariant
contribution from the $(\det\cG)^{-1/2}$ in (\ref{ACTG}) which arose from
the Gaussian integration.

The generalization to 4D is interesting because it separates the
action again into a topological generally covariant action (which describes
the pure gauge configurations) plus an action proportional to $\as$.
We have expressed it in terms of a complicated, albeit local, Lagrangian
that contains a (non-positive) 4D metric and a spin-2 self-dual
tensor. It appears to be different from Lunev's gauge invariant
formulation of 4D $SU(2)$ \cite{Lunv4D}. Restricting to zero instanton
number, the metric becomes conformal. Thus, the sector of the $SU(2)$ theory
with a trivial gauge bundle can be transformed into a theory with
a local Lagrangian with 6 degrees of freedom -- five from the spin-2
self-dual tensor and  an additional degree of freedom from the conformal
metric. We intend to elaborate on this description in a later paper
\cite{INPREP}.

It is an interesting algebraic problem to express the Lagrangian
 (\ref{FFB})
for general $SU(N)$ in terms of local gauge invariant variables in such
a way that is suitable for a large $N$ expansion.

It is also interesting to relate the Lagrangian (\ref{FFB}) to
the Hamiltonian approach of \cite{DZFHJL,BauDZF}.

Finally, for the 3D theory, in a recent work Das and Wadia\cite{DasWad}
have generalized Polyakov's result\cite{Polyak} and
have shown how ``dressed" monopoles generate confinement. It is interesting
to extract from the Lagrangian (\ref{FFB}) that part of the action
that corresponds to just integrating over the collective coordinates
of the monopoles in $\DAu$ in (\ref{ZFUNAG}) and compare to the full
Lagrangian.

\section*{Acknowledgements}
We are very grateful to  D. Kutasov,
and A. B. Zamolodchikov for useful discussions.
We wish to thank D.Z.Freedman for discussing with us his
works, pointing our attention
to the Wu-Yang ambiguity and  familiarizing us with Lunev's result.

\section*{Appendix A: On the cohomology $H^2(SU(N)/U(1)^{N-1})$}
{}From the fibre bundle
\be
\begin{array}{rcc}
U(1)^{N-1}  &  \longrightarrow & SU(N)   \\
            &                  & \downarrow \\
            &                  & SU(N)/U(1)^{N-1}
\end{array}
\ee
We obtain, by standard spectral sequence arguments, a basis
for $H^2(SU(N)/U(1)^{N-1}) = \bfZ^{N-1}$.
Let the $(N-1)$ $U(1)$-s correspond to a Cartan torus, which is generated
by $T^i, i=1,\dots,N-1$ in the Cartan subalgebra.
We choose
\be
T^i =
\left(\begin{array}{ccccc}
     0  & \cdots    &  0        & \cdots  & 0         \\
 \vdots & \ddots    & \vdots    &         & \vdots    \\
     0  & \cdots    & 1_{(i,i)} & \cdots  & 0         \\
 \vdots &           & \vdots    & \ddots  & \vdots    \\
     0  & \cdots    &  0        & \cdots  & 0
      \end{array}
\right)
-\inv{N}\bI
\ee
so that
\be
\tr{T^i T^j} = \dlu{ij}-\inv{N}
\ee
We construct the $i$-th generator of $H^2$ as the Chern class of the
projection
\be
\frac{SU(N)}{W_i\otimes \bfZ_N}
\stackrel{\pi_i}{\longrightarrow} \frac{SU(N)}{U(1)^{N-1}}
\ee
where $W_i$ is generated by the $e^{i T}$-s with $T$ is the
Cartan subalgebra such that $\tr{T T^i} =0$.
To write the Chern class explicitly let
\be
A^{(i)} = \tr{T^i g^{-1}dg},\qquad g\equiv gh\qquad
\mbox{for\ }h\in W_i\otimes \bfZ_N
\ee
$A^{(i)}$ is a one-form field that is well defined for $g\equiv gh$ when
\be
h\in W_i\otimes \bfZ_N
\ee
and transforms as a $U(1)_i$ gauge field for $h\in U(1)_i$.
The two-form $dA^{(i)}$ is the desired Chern class
\be
C^{(i)} = dA^{(i)} =
\tr{T^i (g^{-1}dg) \wedge (g^{-1}dg)}
\ee
which is well defined for $g\equiv gh$ with $h\in U(1)^{N-1}$.
Standard spectral sequence arguments show that the $C^{(i)}$
span $H^2$ and that $H^1 = 0$. Since every map from a two dimensional
CW-complex to another CW-complex can be homotopically pushed
to the two-skeleton of the target complex, we see that the maps $f$
from the torus to $SU(N)/U(1)^{N-1}$ are characterized by
the $N-1$ integer classes of $f^* C^{(i)}$ on the torus.

\section*{Appendix B: On $SU(2)$ gauge invariant expressions}
We are given an expression of the form
\be
R = S^{a,b}_{ij,\u\v}(F) \px{\u}\F{a}{i}\px{\v}\F{b}{j}
+ W^a_{i,\u\v}(F) \px{\u}\px{\v}\F{a}{i}
\label{RGI}
\ee
where $S^{a,b}_{ij,\u\v}(F)$ and $W^a_{i,\u\v}(F)$ are local algebraic
expressions in $\F{a}{i}$ with $a=1,2,3$ an $SU(2)$ index and
$i=1,\dots,K$ (with $K=D(D-1)/2$ for our purposes).
Supposing that $R$ is gauge invariant under local gauge
transformations which transform the $\F{a}{i}$ as
\be
\delta_\h\F{a}{i} = \epu{abc}\h^b\F{c}{i}
\label{FTOF}
\ee
Our goal is to write (\ref{RGI}) in a manifestly invariant form, i.e.
\be
R = \sum_{(\a)}
\varpi_{(\a),\u\v}\px{\u}\varsigma_{(\a)}\px{\v}\varphi_{(\a)}
+\sum_{(\a)}
\varrho_{(\a),\u\v}\px{\u}\px{\v}\vartheta_{(\a)}
\ee
where $\varpi_{(\a),\u\v}$,$\varrho_{(\a)(\b),\u\v}$,
$\vartheta_{(\a)}$ and $\varsigma_{(\a)}$
are gauge invariant algebraic expressions.

By substituting (\ref{FTOF}) in (\ref{RGI}) it is clear that
for each $\u\v$
\be
W^a_{i,\u\v}(F) \frac{d}{dt}\F{a}{i}
\label{WDTFUV}
\ee
(where now we treat $\F{a}{i}(t)$ as depending on a single
parameter $t$) is also gauge invariant.
So our first task will be to write (\ref{WDTFUV}) as
\be
\sum_{(\a)}\widetilde{\varrho}_{(\a),\u\v}\frac{d}{dt}
\widetilde{\vartheta}_{(\a)}
\ee
We then subtract from $R$
\be
\sum_{(\a)}\widetilde{\varrho}_{(\a),\u\v}\px{\u}\px{\v}
\widetilde{\vartheta}_{(\a)}
\ee
The resulting expression does not contain second derivatives and
is still gauge invariant.
Thus the problem reduces to writing invariantly the two separate
expressions:
\bear
W(t)    &\eqdf& W^a_{i}(F)\frac{d}{dt}\F{a}{i}
\label{WDTF}\\
S(x^\u) &\eqdf& S^{a,b}_{ij,\u\v}(F) \px{\u}\F{a}{i}\px{\v}\F{b}{j}
\label{SDUF}
\eear
We will separate the discussion to $K=3$ and $K>3$.

\subsection*{Three field strengths: $K=3$}
In this case the matrix $\F{a}{i}$ is a $3\times 3$ matrix and is
generically invertible. Denote the inverse by $R^a_i$
\bear
R^a_i &=& \inv{2\DD}\epu{abc}\ep{ijk}\F{b}{j}\F{c}{k}  \nn\\
\DD   &=& \inv{6}\epu{abc}\ep{ijk}\F{a}{i}\F{b}{j}\F{c}{k} = \det(F) \nn\\
\F{a}{i} R^a_j &=& \dl{ij} \nn\\
\F{a}{i}\F{b}{i} &=& \dlu{ab} \nn
\eear
We can now write
\bear
W^a_{i} &=& (W^b_i R^b_k)\F{a}{k} \eqdf W_{ik}\F{a}{k}     \\
S^{a,b}_{ij,\u\v} &=& (S^{c,d}_{ij,\u\v} R^c_k R^d_l)
                      \F{a}{k}\F{b}{l}
                      \eqdf S_{ijkl,\u\v}\F{a}{k}\F{b}{l}
\eear
The newly defined quantities $W_{ik},S_{ijkl,\u\v}$ are algebraic
gauge invariants.
Now the gauge invariant expressions  (\ref{WDTF}-\ref{SDUF}) read
\bear
W(t)    &=& W_{ik}(F)\F{a}{k}\frac{d}{dt}\F{a}{i}          \\
S(x^\u) &=& S_{ijkl,\u\v}(F)\F{a}{k}\F{b}{l}\px{\u}\F{a}{i}\px{\v}\F{b}{j} \\
S_{ijkl,\u\v} &=& S_{jilk,\v\u}
\eear
Now we use gauge invariance (and the symmetry of $S$) to obtain
\bear
W_{ik}\epu{abc}\F{a}{k}\F{b}{i} &=& 0    \\
S_{ijkl,\u\v}\epu{abc}\F{a}{k}\F{b}{i} &=& 0
\eear
where in the second equation we used the fact that $\F{b}{l}$ is
(generically) a nonsingular matrix,
and also that $\px{\v}\F{b}{j}$ is generic.
Finally, since
\be
\epu{abc}\F{a}{k}\F{b}{i}\F{c}{j} = \ep{ijk}\DD
\ee
and generically $\DD\ne 0$ we obtain
\bear
W_{ik} &=& W_{ki} \nn\\
S_{ijkl,\u\v} &=& S_{kjil,\u\v}  \nn
\eear
Using
\be
T_{ij} \eqdf \F{a}{i}\F{a}{k}
\ee
we can write
\bear
W(t)    &=& \hlf W_{ik}(F)\frac{d}{dt}(\F{a}{k}\F{a}{i})
         = \hlf W_{ik}(F)\frac{d}{dt}T_{ik}                   \\
S(x^\u) &=& \inv{4} S_{ijkl,\u\v}(F)\px{\u}T_{ik}\px{\v}T_{lj}
\eear
Which is explicitly invariant.

\subsection*{More than three field strengths: $K>3$}
We will limit ourselves to $W(t)$. The other invariant $S$ is
manipulated similarly.
Define the $3\times 3$ matrix
\be
M^{ab} \eqdf \F{a}{i}\F{b}{i}
\ee
It is generically non-singular.
The inverse of a $3\times 3$ matrix $\bM$ is given, by the Cayley-Hamilton
theorem:
\bear
\bM^{-1} &=& \inv{\Lambda} (\bM^2 -\tr{\bM}\bM
        +\hlf (\tr{\bM}^2-\tr{\bM^2})\bI)          \\
\Lambda &=& \det(\bM) =
\inv{3}\tr{\bM^3}-\hlf\tr{\bM}\tr{\bM^2}+\inv{6}\tr{\bM}^3
\eear
and using this we can write
\be
W^a_{i} = W^b_{i}(\bM^{-1})^{bc}\bM^{ca}
          = (W^b_{i}(\bM^{-1})^{bc}\F{c}{k})\F{a}{k}
\ee
So defining  the gauge invariant algebraic expression
\be
W_{ik} \eqdf W^b_{i}(\bM^{-1})^{bc}\F{c}{k}
\ee
we get
\be
W(t) = W_{ik}\F{a}{k}\frac{d}{dt}\F{a}{i}
\ee
Since  $W(t)$ is supposed to be gauge invariant we have
\be
W_{ik}\epu{abc}\F{a}{i}\F{b}{k} = 0
\label{WCF}
\ee
for $c=1,2,3$.
We can decompose $W_{ik}$ into symmetric and anti-symmetric parts
\bear
W^{(A)}_{ik} &\eqdf& \hlf (W_{ik} - W_{ki}) \nn\\
W^{(S)}_{ik} &\eqdf& \hlf (W_{ik} + W_{ki}) \nn
\eear
When we plug the symmetric part into $W(t)$ we get the gauge invariant
expression
\be
W^{(S)}(t) = \hlf W^{(S)}_{ik}(F)\frac{d}{dt}(\F{a}{k}\F{a}{i})
         = \hlf W_{ik}(F)\frac{d}{dt}T_{ik}
\ee
So from now on we will assume that $W_{ik}$ is anti-symmetric,
$W_{ik} = -W_{ki}$.
Defining
\be
R^c_{ik} \eqdf \epu{abc}\F{a}{i}\F{b}{k}
\ee
which is antisymmetric in $ik$.
(\ref{WCF}) expresses the fact that the $K(K-1)/2$ vector
$W_{ij}$ is orthogonal to the three vectors $R^c_{ik}$.
We need a projection operator on the space that is orthogonal
to $R^c_{ik}$. Such a projection operator is given as follows.
Suppose $W_{ij}$ is anti-symmetric but not necessarily satisfying
(\ref{WCF}), then
\bear
V_{ij} &=& W_{ij} - 3\mu^{-1} W_{kl}K_{klm}K_{ijm} \\
\mu    &=& K_{klm}K_{klm}
\eear
where
\bear
T_{12} &\eqdf& \dG{a}{1}\dG{a}{2}                      \label{TINVAR}\\
K_{123} &\eqdf& \epu{def}\dG{d}{1}\dG{e}{2}\dG{f}{3}   \label{KINVAR}
\eear
If $W_{ij}$ satisfies (\ref{WCF}) then $V_{ij}=W_{ij}$.
Furthermore, for any $W_{ij}$, not necessarily satisfying (\ref{WCF}) we
have $V_{ik}\epu{abc}\F{a}{i}\F{b}{k} = 0$.
Thus it has to be that for any $W_{ij}$ the expression
$V_{ik}\F{a}{k}\frac{d}{dt}\F{a}{i}$ can be written in a manifestly
invariant way. This is indeed so. Define
\bear
U_{ij} &\eqdf& T_{ik}T_{kj} = ({\bf T}^2)_{ij}    \nn\\
G_{ij} &\eqdf& T_{ik}T_{kl}T_{lj} = ({\bf T}^3)_{ij}    \nn\\
T      &\eqdf& T_{ii} = \tr{\bf T}  \nn\\
U      &\eqdf& U_{ii} = \tr{{\bf T}^2}  \nn\\
G      &\eqdf& G_{ii} = \tr{{\bf T}^3}  \nn\\
\mu    &\eqdf& T^3-3UT+2G = 6\det{M^{ab}}  \nn
\eear
\be
P_{ij} \eqdf 6\mu^{-1} (G_{ij}-TU_{ij} +\hlf(T^2-U)T_{ij})
\ee
$P_{ij}$ is a projection operator and satisfies
\be
P_{ij}\F{a}{j} = \F{a}{j}
\ee
Now we can write
\bear
V_{ij}\F{a}{j}\frac{d}{dt}\F{a}{i} &=&
(W_{ij} - 3\mu^{-1} W_{kl}K_{klm}K_{ijm})\F{a}{i}\frac{d}{dt}\F{a}{j} \nn\\
= -2P_{ik}W_{ij}\frac{d}{dt}T_{kj}
&+&6\mu^{-1}(T_{lm}K_{mik}-\hlf T K_{lik})W_{ij}\frac{d}{dt}K_{klj}
\nn\\&& \label{VWINV}
\eear
which is written in terms of local invariant objects.

\section*{Appendix C: Expressing $\Lag$ in terms
                      of $W_{\u\v\s\t}$ and $g_{\u\v}$}

In appendix (B) we described in general how to write $SU(N)$ Lagrangians
like (\ref{FFB}) in gauge invariant variables. We saw that the $D=3$
case is simpler because the field strength $\GG{a}{\u\v}$
can be thought of as a $3\times 3$ matrix (three values of $a$, and
three values of $\u\v$). In higher dimensions, $D>3$ this is not the
case and we had more complications.
However, for $D=4$, we have seen in section (5), that when passing
to the special metric (\ref{GINVAR}), the field strength becomes
self-dual. Thus, given the metric, there are only three linearly
independent $\u\v$-s, and $\Fd{a}{\u\v}$ is effectively a $3\times 3$
matrix.

We shall now describe in detail how the invariant Lagrangian is obtained.
We start with the topological part of (\ref{FFB}):
\bear
\Lag_{topo} &=&
2\dG{a}{\u\v}\VG{ab}{\v\t}\px{\u}\WG{bc}{\t\s}
       \VG{cd}{\s\r}\px{\a}\dG{d}{\a\r}
\nn\\
&-&
2\dG{a}{\u\v}\VG{ab}{\v\t}\px{\u}\px{\s}\dG{b}{\s\t}
-\VG{cd}{\v\r}\px{\a}\dG{c}{\a\v}\px{\b}\dG{d}{\b\r}
\nn
\eear
Since the topological part is independent of the metric
we pick our special induced metric (\ref{GINVAR}) and write
\bear
\Lag_{topo} &=&
2\Fu{a}{\u\v}\VG{ab}{\v\t}\epu{bce}\Kv_{\u}\Fu{e}{\t\s}
       \VG{cd}{\s\r}\Kv_{\a}\Fu{d}{\a\r}
\nn\\
&-&
2\Fu{a}{\u\v}\VG{ab}{\v\t}\Kv_{\u}\Kv_{\s}\Fu{b}{\s\t}
-\VG{cd}{\v\r}\Kv_{\a}\Fu{c}{\a\v}\Kv_{\b}\Fu{d}{\b\r}
\nn
\eear
where $\Kv_\u$ is the covariant derivative with respect to that
metric.
We substitute our covariant expression for $\VG{ab}{\u\v}$ and obtain
\bear
\Lag_{topo} &=& \Lag_1 +\Lag_2 +\Lag_3 \nn\\
\Lag_1 &=&
\hlf\Fdu{c}{\v}{\g}\Fd{d}{\g\r} \Kv_\a\Fu{c}{\a\v}\Kv_\b\Fu{d}{\b\r}
+\inv{4}\Fu{c}{\a\b}\Fd{d}{\a\b}g_{\v\r} \Kv_\a\Fu{c}{\a\v}\Kv_\b\Fu{d}{\b\r}
\nn\\
\Lag_2 &=&
\Fdu{a}{\v}{\g}\Fd{b}{\g\t}\Fu{a}{\u\v}\Kv_\u\Kv_\r\Fu{b}{\r\t}
+\hlf\Fud{a}{\u}{\t}\Fu{a}{\a\b}\Fd{b}{\a\b}\Kv_\u\Kv_\s\Fu{b}{\s\t}
\nn\\
\Lag_3 &=&
\hlf\Fu{a}{\u\v}\Fdu{a}{\v}{\g}\Fd{b}{\g\t}
\epu{bce}\Kv_\u\Fu{e}{\t\s}\Fdu{c}{\s}{\g}\Fd{d}{\g\r}\Kv_\a\Fu{d}{\a\r}
\nn\\ &+&
\inv{4}\Fu{a}{\u\v}\Fdu{a}{\v}{\g}\Fd{b}{\g\t}
\epu{bce}\Kv_\u\Fud{e}{\t}{\s}\Fu{c}{\g\d}\Fd{d}{\g\d}\Kv_\a\Fu{d}{\a\s}
\nn\\
&+&
\inv{4}\Fud{a}{\u}{\t}\Fu{a}{\a\b}\Fd{b}{\a\b}
\epu{bce}\Kv_\u\Fu{e}{\t\s}\Fdu{c}{\s}{\g}\Fd{d}{\g\r}\Kv_\a\Fu{d}{\a\r}
\nn\\
&+&
\inv{8}\Fud{a}{\u}{\t}\Fu{a}{\a\b}\Fd{b}{\a\b}
\epu{bce}\Kv_\u\Fud{e}{\t}{\s}\Fu{c}{\g\d}\Fd{d}{\g\d}\Kv_\a\Fu{d}{\a\s}
\nn
\eear

It is tedious though straight-forward to check the identity
\bear
\Fdu{a}{\u}{\g}\Fd{a}{\g\v} &=& -3\Phi g_{\u\v} \nn
\eear
We use it to express $\Lag_2$ as
\bear
\Lag_2 &=& \Lag_{2}'
+\frac{3}{2}\Phi\Kv^\u\Kv_\u\Phi - R\Phi^2+\inv{16}R\WW
\nn\\
&-&\inv{4}\Phi\Rd{\u\v\s\t}\Wu{\u\v\s\t}
-\inv{8}\Rd{\u\v\s\t}\Wud{\s\t}{\a\b}\Wu{\a\b\s\t}
\nn\\
&+&\inv{4}\Wud{\a\b\u}{\t}\Kv_\u\Kv_\s\Wd{\a\b\s\t}
+\inv{4}\Wud{\a\b\s}{\t}\Kv^\u\Kv_\s\Wd{\a\b\u\t}
-\inv{4}\Wu{\a\b\g\d}\Kv^\u\Kv_\u\Wd{\a\b\g\d}
\nn\\
\Lag_{2}' &=&
-\hlf\Phi\Kv_\u\Fud{b}{\u}{\t}\Kv_\s\Fu{b}{\s\t}
-\hlf\Phi\Kv_\s\Fud{b}{\u}{\t}\Kv_\u\Fu{b}{\s\t}
\nn\\
&-&\hlf\Wud{\a\b\u}{\t}\Kv_\u\Fd{b}{\a\b}\Kv_\s\Fu{b}{\s\t}
-\hlf\Wud{\a\b\u}{\t}\Kv_\s\Fd{b}{\a\b}\Kv_\u\Fu{b}{\s\t}
\nn\\
&+&
\hlf\Wu{\a\b\g\d}\Kv^\u\Fd{b}{\a\b}\Kv_\u\Fd{b}{\g\d}
\nn
\eear

where $R$ is the curvature:
\bear
R_{\a\b\g\d} &=& g_{\d\u}(\px{\a}\Gamma^\u_{\b\g} - \px{\b}\Gamma^\u_{\a\g}
+ \Gamma^\u_{\a\s}\Gamma^\s_{\b\g} - \Gamma^\u_{\b\s}\Gamma^\s_{\a\g})
\nn\\
R_{\a\b} &=& g^{\g\d} R_{\a\g\b\d}
\nn\\
R &=& g^{\a\b} R_{\a\b}
\nn
\eear
Now we repeat the arguments of appendix (B) for the $D=3$ case
to argue that if an expression of the form
\be
\uus{\u\v}{\a\b}{\kp}{\s\t}{\g\d}{\r}
\Fu{a}{\u\v}\Kv^\kp\Fu{a}{\a\b}
\Fu{b}{\s\t}\Kv^\r\Fu{b}{\g\d}
\ee
is gauge invariant, it can be written as
\be
\inv{4} \uus{\u\v}{\a\b}{\kp}{\s\t}{\g\d}{\r}
\Kv^\kp\SWu{\u\v\a\b} \Kv^\r\SWu{\s\t\g\d}
\ee
where
\bear
\Hd{\u\v\s\t} &\eqdf& g_{\u\s}g_{\v\t} - g_{\u\t}g_{\v\s}
                      +\sqrt{g} \ep{\u\v\s\t} \nn\\
\Phi &=& \inv{24\sqrt{g}}\epu{\u\v\s\t}\Fd{a}{\u\v}\Fd{a}{\s\t} \nn\\
\SWd{\u\v\s\t}  &\eqdf& \Wd{\u\v\s\t} +\Phi \Hd{\u\v\s\t} \nn
\eear
In order to express our Lagrangian in a suitable form
we need two more identities that can all be induced from
the $3\times 3$ nature of the matrices involved (see e.g.
appendix (D)).
For $\Lag_3$ we need
\be
\epu{abc}\Fd{b}{\u\t}\Fd{c}{\s\g}
  = \pps{\u\t}{\s\g}{\a\b}\Fu{a}{\a\b}
\ee
with
\bear
&& \pps{\u\t}{\s\g}{\a\b} = \nn\\
&& \inv{16}\WWd{\a\b\t\g} g_{\u\s}
  -\inv{16}\WWd{\a\b\t\s} g_{\u\g}
  -\inv{16}\WWd{\a\b\u\g} g_{\t\s}
  +\inv{16}\WWd{\a\b\u\s} g_{\t\g}
\nn\\
&-&\inv{4}\Phi\Wd{\a\b\t\g} g_{\u\s}
  +\inv{4}\Phi\Wd{\a\b\t\s} g_{\u\g}
  +\inv{4}\Phi\Wd{\a\b\u\g} g_{\t\s}
  -\inv{4}\Phi\Wd{\a\b\u\s} g_{\t\g}
\nn\\
&+&(\inv{4}\Phi^2 - \inv{128}\WW )(
   \Hd{\a\b\t\g} g_{\u\s}
  -\Hd{\a\b\t\s} g_{\u\g}
  -\Hd{\a\b\u\g} g_{\t\s}
  +\Hd{\a\b\u\s} g_{\t\g})
\nn
\eear
\bear
\WWd{\u\v\s\t} &\eqdf&  \Wdu{\u\v}{\a\b}\Wd{\a\b\s\t} \nn\\
\WW            &\eqdf&  \Wu{\u\v\s\t}\Wd{\u\v\s\t} \nn
\eear
For $\Lag_2$ we need
\be
\dlu{ab} = \vvs{\u\v\s\t}\Fu{a}{\u\v}\Fu{b}{\s\t}
\ee
\be
\vvs{\u\v\s\t} \eqdf
\inv{64}\WWd{\u\v\s\t} -\inv{16}\Phi\Wd{\u\v\s\t} +
\inv{16}(\Phi^2 -\inv{32}\WW)\Hd{\u\v\s\t}
\ee
Putting everything together we get the expression fo the
Lagrangian:
\bear
&&\frac{3}{8}\Phi\Kv^\u\Kv_\u\Phi -R\Phi^2+\inv{16}R\WW
-\inv{4}\Phi\Rd{\u\v\s\t}\Wu{\u\v\s\t}
-\inv{8}\Rd{\u\v\s\t}\Wud{\s\t}{\a\b}\Wu{\a\b\s\t}
\nn\\
&+&\inv{4}\Wud{\a\b\u}{\t}\Kv_\u\Kv_\s\Wd{\a\b\s\t}
+\inv{4}\Wud{\a\b\s}{\t}\Kv^\u\Kv_\s\Wd{\a\b\u\t}
-\inv{4}\Wu{\a\b\g\d}\Kv^\u\Kv_\u\Wd{\a\b\g\d}
\nn\\
&+&\inv{16}\Kv_\a\SWu{\a\v\g\d} \Kv^\b\SWd{\b\v\g\d}
-\inv{8}\Phi g_{\g\d}\vvs{\u\v\s\t}\Kv_\a\SWu{\a\g\u\v}\Kv_\b\SWu{\b\d\s\t}
\nn\\
&-&\inv{8}\Phi g_{\g\d}\vvs{\u\v\s\t}\Kv_\b\SWu{\a\g\u\v}\Kv_\a\SWu{\b\d\s\t}
-\inv{8}\Wud{\a\b\u}{\t}\vvs{\g\d\v\s}
                      \Kv_\u\SWud{\g\d}{\a\b}\Kv_\r\SWu{\r\t\v\s}
\nn\\
&-&\inv{8}\Wud{\a\b\u}{\t}\vvs{\g\d\v\s}
                      \Kv_\r\SWud{\g\d}{\a\b}\Kv_\u\SWu{\r\t\v\s}
+\inv{8}\Wu{\a\b\g\d}\vvs{\u\v\s\t}\Kv^\r\SWud{\u\v}{\a\b}
                                            \Kv_\r\SWud{\s\t}{\g\d}
\nn\\
&-&\frac{3}{8}\Phi\pps{\u\t}{\s\a}{\b\d}\Kv^\u\SWu{\b\d\t\s}\Kv^\a\Phi
-\inv{16}\Phi\pps{\u\t}{\s\g}{\b\d}
         \Kv^\u\SWu{\b\d\t\r}\Kv_\a\SWud{\s\g\a}{\r}
\nn\\
&+&\frac{3}{16}\Wud{\a\b\u}{\t}\pps{\a\b}{\s\r}{\g\d}
                              \Kv_\u\SWu{\g\d\t\s}\Kv^\r\Phi
+\inv{32}\Wud{\a\b\u}{\t}\pps{\a\b}{\s\r}{\g\d}
             \Kv_\u\SWu{\g\d\t\v}\Kv_\kp\SWud{\s\r\kp}{\v}
\nn
\eear
with

\section*{Appendix D: Expressing $\Phi$ in terms of $W_{\u\v\s\t}$}
$W_{\u\v\s\t}$ and $\Phi$ are defined as
\bear
W_{\u\v\s\t} &=&
\Fd{a}{\u\v}\Fd{a}{\s\t} -
\inv{24\sqrt{g}}\epu{\a\b\g\d}\Fd{a}{\a\b}\Fd{a}{\g\d}
(g_{\u\s}g_{\v\t}-g_{\u\t}g_{\v\s} +\sqrt{g} \ep{\u\v\s\t})
\nn\\
\Phi &=& \inv{24\sqrt{g}} \epu{\u\v\s\t}\Fd{a}{\u\v}\Fd{a}{\s\t}
\nn
\eear
Define the $3\times 3$ matrix
\be
\bM^{ab} \eqdf \inv{8\sqrt{g}} \epu{\u\v\s\t}\Fd{a}{\u\v}\Fd{b}{\s\t}
\ee
It is easy to see that
\be
\det\bM = \frac{\DD}{8\sqrt{g}^3} = 1
\ee
Let $\lam_1,\lam_2,\lam_3$ be the three eigen-values of $\bM$.
Then
\bear
\Phi &=&  \inv{3}(\lam_1 +\lam_2 +\lam_3) \nn\\
1 &=& \lam_1\lam_2\lam_3  \nn
\eear
thus we need two more relations among $\lam_1,\lam_2,\lam_3$ to determine
$\Phi$.
Using the self-duality of $W$ we have
\be
\inv{16}W^{\u\v\s\t} W_{\u\v\s\t}
=(\inv{4\sqrt{g}})^2\epu{\u\v\a\b}\epu{\s\t\g\d}W_{\u\v\s\t} W_{\a\b\g\d}
= \tr{\bM^2}-3\Phi^2
\ee
 From a similar equation for $\tr{\bM^3}$
\be
\tr{\bM^3} = \inv{64}W^{\u\v\s\t} {W_{\u\v}}^{\a\b}  W_{\a\b\s\t}
+\frac{3}{16}\Phi W^{\u\v\s\t} W_{\u\v\s\t} +3\Phi^2
\ee
we find that  $\Phi$ is the solution of the cubic equation
\be
\Phi^3 - A\Phi -B =0
\ee
with
\bear
A &=&  \inv{32} W^{\u\v\s\t} W_{\u\v\s\t} \nn\\
B &=&  1- \inv{192} W^{\u\v\s\t} {W_{\u\v}}^{\a\b}  W_{\a\b\s\t}\nn
\eear

\end{document}